# Understanding differences of the OA uptake within the Germany university landscape (2010-2020) – Part 2: repository-provided OA

Niels Taubert[1] (ORCID: 0000-0002-2357-2648), Anne Hobert[2] (0000-0003-2429-2995), Najko Jahn[2] (0000-0001-5105-1463), Andre Bruns[1] (0000-0002-2976-0826), Elham Iravani[1] (0000-0003-1961-2130)

[1] Institute for Interdisciplinary Studies of Science (I²SoS), Bielefeld University, Germany
[2] Göttingen State and University Library, University of Göttingen, Germany

Correspondence: niels.taubert@uni-bielefeld.de

## Abstract

This study investigates the determinants for the uptake of institutional and subject repository Open Access (OA) in the university landscape of Germany and considers three factors: The disciplinary profile of universities, their OA infrastructures and services and large transformative agreements. The uptake of OA as well as the determinants are measured by combining several data sources (incl. Web of Science, Unpaywall, an authority file of standardised German affiliation information, the ISSN-Gold-OA 4.0 list, and lists of publications covered by transformative agreements). For universities' OA infrastructures and services, a structured data collection was created by harvesting different sources of information and by manual online search. To determine the explanatory power of the different factors, a series of regression analyses was performed for different periods and for both institutional as well as subject repository OA. As a result of the regression analyses, the most determining factor for the explanation of differences in the uptake of both repository OA-types turned out to be the disciplinary profile, whereas all variables that capture local infrastructural support and services for OA turned out to be non-significant. The outcome of the regression analyses is contextualised by an interview study conducted with 20 OA officers of German universities. The contextualisation provides hints that the original function of institutional repositories, offering a channel for secondary publishing is vanishing, while a new function of aggregation of metadata and full texts is becoming of increasing importance.

## Declarations

Funding: This work was supported by the German Federal Ministry of Education and Research within the funding stream "Quantitative research on the science sector", project OAUNI (grant numbers 01PU17023A and 01PU17023B).

Conflicts of interest: The authors declare that they have no conflict of interest.

# 1. Introduction

Open Access (OA) to scholarly literature has become an important and prominent way of communicating research (Robinson-Garcia et al., 2020). Recent studies show that there is a general trend towards a growing proportion of OA articles worldwide (Piwowar et al. 2018) and with exceptions also for individual countries (Bosman & Kramer, 2018), disciplines (Bosman & Kramer, 2018) and institutions (Huang et al., 2020, Hobert et al., 2021). However, this trend has multiple facets. Under the broad categories of journal-provided (or 'Gold') and repository-provided (or 'Green') OA as suggested by Suber (2012), various subtypes have evolved (Taubert et al., 2019). This article focuses on repository-provided OA, asks for possible determinants for differences in the uptake on the level of institutions and takes the German university landscape as an example. In this publication model, repositories act as a dissemination channel besides journals and papers may be deposited before or after acceptance in a journal. Furthermore, the type of repository on which a document is deposited can differ. The two major types are subject repositories (SR), which address a certain discipline or scientific field, and institutional repositories (IR), which refer to the members of an institution as the intended user group.

On the level of research institutions, the transition towards OA is complex as many factors may be relevant for the volume of OA that is achieved. In this study, three different types of factors are investigated: first, there is evidence that repository-provided OA has been adapted by different disciplines to a different extent and at different points in time (Bosman & Kramer, 2018; Severin et al., 2020). Therefore, the OA-affinity of the disciplinary profiles of the institution is considered as one factor. Second, there is evidence at hand that the extent to which German universities provide OA-supporting infrastructures and services differ (Kindling et al., 2021). Thus, the available infrastructural support on the level of institutions is included as a second type of factor. Third, repository-provided OA is not only achieved by actions of individual scientists that deposit their article (self-archiving) but also depends on the university's strategy that is applied for the aggregation of content from different sources (Poynder, 2016). The availability of such content depends at least to some extent on initiatives that influence the whole university landscape. In our attempt to explain repository-provided OA shares, the availability of OA content for aggregation via project DEAL and subject repositories is incorporated as a third possible factor.

The article is organised as follows: in a first step, a brief overview about the development of the repository landscape is given for both subject and institutional repositories (Section 2). Given that SR and IR evolved at different points in time and serve different purposes, separate hypotheses are developed for each repository type (Section 3). The methods, the operationalisation of the different factors and the data collection is described in in Section 4, followed by descriptive statistic and statistical tests of the hypotheses in Section 5. The results of the analysis are contextualised with evidence drawn from 20 interviews with OA administrators from German universities (Section 6). The concluding section (Section 7) summarises the most important results.

# 2. The development of the repository landscape

The repository landscape is older than the OA movement as first experiments with preprint servers date back as early as 1991 both for physics (Ginsparg, 1994, 2011) and mathematics (Jackson, 2002). Interestingly, the initiative for the creation of such SR emanated from initiatives of the community of scholars in the two fields. Similar activities can be found in the sharing of economic literature through the REPEC network. RePEc started in 1997 and offers free online access to published and unpublished works. PubMed Central, a full text repository for articles published in medical sciences journals, was founded a few years later in 2000 (Marschall, 1999; Roberts, 2001) but differs in the way in which content is acquired. While preprint servers are fed by members of the scientific community who aim to speed up dissemination of research for different reasons or gather feedback from colleagues (Taubert, 2021), PubMed Central archives mainly already published articles in cooperation with a number of well-established life science journals. Therefore, the majority of articles are peer reviewed and not provided by authors but by journals. In more recent years, SR were populated in other hitherto unaffected disciplines by the Center for Open Science at the University of Maryland, which runs SocArXiv[1]. With the COVID-19 health crisis, discipline-specific preprint servers have proliferated to disseminate research in fields where such a practice was not previously well established (Fraser et al., 2021). Examples include medRxiv[2] for health sciences and medicine and bioRxiv[3] for the biological sciences, both run by Cold Spring Harbor Laboratory.

Before the creation of the first institutional repositories (IR) from 2002 onwards (Lynch, 2003), SR had already been around for a decade. From their very beginning, the aim of IRs was not as clear as for SR. Part of their proponents took SR as a role model and defined their function as providing free electronic access to publications that were published in traditional, toll access publications channels – journals in first place (Crow, 2002; Harnad, 2013). In doing so, they were understood as a reaction to the journal crisis and a means for the disruption of the journal system towards OA. A second approach defined the role of IR as a publication channel for the dissemination of all types of intellectual products from research and teaching (Lynch, 2003) that were usually excluded from the traditional publication channels. Such products include grey literature, technical reports, working papers, theses, dissertations, video and audio recordings, research instruments and protocols as well as datasets and software (Kennison et al. 2013). However, both visions of IR have the common ground that faculty should get involved in the usage of the IR by the provision of content via self-archiving.

Regarding the development of the repository landscape, a number of studies show that a wave of IR was established in the middle of the first decade of the millennium in many countries (Pinfield et al., 2014). In 2005, Australia, Norway, the Netherlands, and Germany already showed large numbers of IR compared with the number of universities, leading to the assumption that a good provision of such services was already in place (van Westrienen & Lynch, 2005). A strong development towards a diversified institutional repository landscape was also observed in the same year for the US (Lynch & Lippincott, 2005). Since 2011, Pinfield et al. (2014) report a plateaued growth for the UK and the Netherlands, indicating a saturation, while the growth of Germany`s repository landscape tended to be slower and more sustained during that time. After a critical discussion of the shortcomings of the respective OA evidence systems (OpenDOAR and ROAR), Arlitsch & Grant (2018) estimate the number of repositories world-wide to be as large as 3,000-3,500 for 2018. Since then, the growth in numbers seems to continue. At the time of writing, OpenDOAR lists 5,995 repositories, of which a small minority of only 375 are SR while the large majority of 5,311 are IR.

---

[1]    https://osf.io/preprints/socarxiv (accessed August 15, 2023).
[2]    https://www.medrxiv.org/ (accessed August 15, 2023).
[3]    https://www.biorxiv.org/ (accessed August 15, 2023).

Besides the sheer quantity of the repositories, some studies are interested in the role that IR and SR play in the communication system of science and we would like to mention four results here: first, there is some evidence of disciplinary differences in the adaption of repository-provided OA (Kling & McKim, 2000; Björk et al., 2014; Archambault et al., 2014; Martín-Martín et al., 2018; Pölönen et al. 2020). Second, such differences can be found for SR but are also observed for IR (Pinfield et al., 2014). Third, whenever scientists are motivated internally to self-archive, SR are often preferred to IR (Spezi et al., 2012). In contrast, when self-archiving is mandated, IR are used more often for the deposition of articles. Fourth, scholars who are familiar with self-archiving in a SR do not necessarily self-archive on IR. "Instead, evidence reveals that when an article has been presented in one repository, the author(s) will be hesitant to make it repeatedly available in a second repository" (Xia 2008, p. 494). In other words, the result suggests a rivalry between SR and IR for authors that self-archive their publications.

The last aspect we would like to mention refers to IR only: the review of the literature reveals that the aim of IR was not only controversial at the beginning but was also altered and redefined along the way of its development. This dynamic was in part driven by the development of web technology and the availability of data, and in part the result of changing priorities of university management. However, one important condition for the continual redefinition of aims are complaints about the low usage of IR for the deposition of content since their origins (Westrienen & Lynch, 2005; Xia 2008; Nicholas et al., 2012; Arlitsch & Grant, 2018; Novak & Day, 2018). The first and obvious reaction to low usage was to sustain the aim of IR to provide free access to electronic content (however defined) but to attribute the responsibility for the deposition in a new way. Even though self-archiving is still an option in many IR, the content was more and more uploaded by library staff on behalf of the scientists. In other words, self-archiving was increasingly replaced by 'mediated archiving' (Xia, 2008). According to an interview study conducted in 2011, the majority of faculty from medical and life sciences as well as authors from social sciences and humanities report that someone else deposited their publication in an IR (Spezi et al., 2012).

A second and more far reaching step in the direction of a re-definition of aims is undertaken when IR-provided OA is linked to research evaluation. In the prominent case of the University of Liège, OA is mandated and the IR is used for the provision of data on which internal research evaluation (for tenure and grant proposals) is based (Rentier & Thirion, 2011). In this case, the requirements of research evaluation are piggybacked on the original aim of providing access to faculties' intellectual work.[4] The confutation of IR with (Current) Research Information Systems as suggested by Dempsey (2014) and Plutchak & Moore (2017) points in the same direction. The collection, enrichment, and synchronisation of metadata like the information about the organisational structure of the institution, ORCID identifiers for authors, Altmetrics, and citation information do not support free access to content in the first place but serve the need for monitoring, evaluation, and reporting of the university management (Novak & Day; Zervas et al., 2019). It is a matter of fact that such an aim does not necessarily need full texts but a collection of valid and well curated metadata to be fulfilled.

# 3. Research question and hypotheses

The overall aim of this article is to answer the question as to what extent three types of factors determine the repository-provided OA share of German universities. These factors are the disciplinary profile of the universities, their local infrastructures and services and transformative agreements. Such agreements are in the first place instruments to foster OA publications in journals (Taubert et al. 2023) but also lead to the proliferation of a large amount of OA content that can be aggregated in repositories. Before explaining the three factors in more detail, the two repository-provided types SR OA and IR OA shall be defined.

- *Subject repository OA (SR OA)* identifies articles for which a version is available on a repository that is used by a certain discipline or field (e.g. arXiv, RePEc, PubMed Central, medRxiv, bioRxiv, and SocArXiv). Such versions may be preprints (i.e. a version of an article that has been made available before acceptance for publication in a journal) or postprints (i.e. a version of an article that has been made available after acceptance for publication in a journal).
- *Institutional repository OA (IR OA)* identifies articles for which a version is available on a repository run by a university or a research organisation. Again, such versions can include pre- as well as postprints.

*Disciplinary profile*: The first type of factor that may determine the OA profile of a university is its disciplinary profile or, in other words, the specific mixture of scholarly outputs in different disciplines, specialties and fields. It is generally known that disciplines share specific publication cultures, communicate via a set of field-specific publication media that are subject to generalised quality attributions of the respective scientific community on which a reputation pyramid of the publication media is based, as well as convictions about standards that publishable research has to meet together with mechanisms that apply such criteria to select contributions for publication (e.g., peer review). Another component are field-specific routines or practices on how to deal with scientific information that is made available via different channels. Publication cultures also comprise attitudes towards different OA-types resulting in differences regarding the OA-affinity (Zhu 2017, Dalton et al., 2020) also with respect to repository-provided OA (Kling & McKim, 2000; Björk et al., 2014, Pinfield et al., 2014, Pölönen et al. 2020).

---

[4]   The amalgamation of OA with research assessment also happens on the country-level. See as an example the UK research excellence framework as described in Ten Holter (2020).

*Local OA infrastructures and services*: a second type of factor that may be relevant for the uptake of OA are infrastructures that are accompanied by services and staff and provided locally by universities. The rationale behind such infrastructure is to offer means to publish research OA and to reduce the related efforts for scientists. With regard to repository-provided OA, local infrastructures and services include IR for depositing (and aggregating) research. Moreover, some universities introduced positions like OA officers that should support scientists, OA websites that inform about the topic as well as OA events and training activities that aim to raise the attention and to teach the necessary competencies. With respect to rules and regulations that support OA, the German situation is special: given that freedom of science is guaranteed by the German constitution (Art. 5 III Grundgesetz), and given that the publication of research is protected by this right, there are no mechanisms of strong top-down regulations on the level of universities like, for example, mandates that enforce repository-provided OA.[5] However, a number of universities have given themselves an OA policy, which expresses the university leaders' support and 'encourages' scientists of the institution to make their publications freely available online. For the German university landscape, mapping instruments (Kindling et al., 2021)[6] show that there are remarkable differences of OA infrastructures and services at different locations.

*Transformative agreements*: in recent years, transformative agreements have been introduced, and the probably most impactful contracts are those that were negotiated between large publishing houses and project DEAL[7]. Given that they operate on an 'all-in-principle' of nearly all public research institutions to date, the contracts can be regarded as a central coordination mechanism that affects the entire German research system. Such instruments are in the first place means to turn publications at the publishers' venue into OA (Haucap et al., 2021) as they make all publications from a member institution in the portfolio of a publisher freely available online while offering access to the whole content of a defined set of journals from a publisher to member institutions. Such contracts create a large stock of publications that are OA and hence can also be used for the aggregation of content in repositories. Therefore, transformative agreements may also affect the OA share of a university that is available via repositories.

## Subject repository OA

After having explained the three factors in more detail, we formulate the following hypotheses regarding their influence on the subject repository OA share of universities.

---

[5]  The only mandate in Germany that requires the archiving of full texts of publications on a repository was adopted by the University of Konstanz in 2015. The question whether or not the mandate violates the freedom of science guaranteed by the German constitution is subject of a lawsuit (Hartmann, 2017).

[6]  https://open-access.network/en/services/oaatlas (accessed August 15, 2023).

[7]  https://www.projekt-deal.de/about-deal/ (accessed August 15, 2023).

*SR-1: Hypothesis on the influence of the disciplinary profile[8]*

H$_1$: Universities with a disciplinary profile that shows a strong affinity towards SR OA have a larger SR OA share than universities with a weaker disciplinary affinity towards SR OA.

*H$_0$*: Universities with a disciplinary profile that shows a strong affinity towards SR OA have a smaller (or equal) SR OA share than universities with a weaker disciplinary affinity towards SR OA.

*SR-2: Hypothesis of the impact of institutional OA-policies[9]*

H$_1$: Universities with an OA policy have a larger SR OA share than universities without an OA policy.

H$_0$: Universities with an OA policy have a smaller (or equal) SR OA share than universities without an OA policy.

*SR-3: Hypothesis on the impact of OA officers, OA information and OA-activities[10]*

*H$_1$:* Universities with an OA officer, websites with OA (rights) information and/or OA training activities and events have a larger SR OA share than universities without such OA-supporting provisions.

*H$_0$*: Universities with an OA officer, websites with OA (rights) information and/or OA training activities and events have a smaller (or equal) SR OA share than universities without such OA-supporting provisions.

*SR-4: Hypothesis on the impact of transformative agreements[11]*

H$_1$: The larger the share of the publication output covered by transformative agreements, the smaller is the SR OA share of German universities.

H$_0$: The larger the share of the publication output covered by transformative agreements, the larger (or equal) is the SR OA share of German universities.

## Institutional repository OA

The following six hypotheses are inspired by the overriding assumption that the IR OA share is predominantly determined by organisational factors and, in particular, infrastructural support.

*IR-1: Hypothesis on the influence of the disciplinary profile[12]*

*H$_1$:* Universities with a disciplinary profile that shows a strong affinity towards IR OA have a larger IR OA share than universities with a weaker disciplinary affinity towards IR OA.

*H$_0$*: Universities with a disciplinary profile that shows a strong affinity towards IR OA have a smaller (or equal) IR OA share than universities with a weaker disciplinary affinity towards IR OA.

---

[8]   The hypothesis is supported by studies that report differences in the uptake of subject repositories in different disciplines and fields (Kling & McKim, 2000; Björk et al., 2014; Archambault et al., 2014; Martín-Martín et al., 2018; Pölönen et al. 2020).

[9]   For the compliance to OA policies and mandates, see Picarra & Swan (2015); Lovett et al. (2017); and Herrmannova et al. (2019).

[10]  In the statistical model, the variables that are summarised in this hypothesis are tested individually.

[11]  Such 'avoidance of double work'-hypothesis is inspired by Xia (2008).

[12]  The hypothesis is inspired by evidence of disciplinary differences in the usage of IR OA (Spezi et al. 2012; Pinfield et al. 2014).

*IR-2: Hypothesis on an infrastructural requirement[13]*

$H_1$: Universities that provide an IR for self-archiving have a larger IR OA share than universities that do not provide such a repository.

$H_0$: Universities that provide an IR for self-archiving have a smaller (or equal) IR OA share than universities that do not provide such a repository.

*IR-3 Hypothesis on OA policies[14]*

$H_1$: Universities with an OA policy have a larger IR OA share than universities that do not have such a policy.

$H_0$: Universities with an OA policy have a smaller (or equal) IR OA share than universities that do not have such a policy.

*IR-4: Hypothesis on the impact of OA officers, OA information and OA-activities [15]*

$H_1$: Universities with an OA officer, websites with OA (rights) information and/or OA training activities and events have a larger IR OA share than universities without such OA-supporting provisions.

$H_0$: Universities with an OA officer, websites with OA (rights) information and/or OA training activities and events have a smaller or equal IR OA share than universities without such OA-supporting provisions.

*IR-5: Hypothesis on the aggregation of OA publications covered by transformative agreements[16]*

$H_1$: The larger the share of the publication output of a university covered by transformative agreements, the larger is the IR OA share (as university libraries find more OA publications from transformative agreements that can be aggregated in their IR).

$H_0$: The larger the share of the publication output of a university covered by transformative agreements, the smaller (or equal) is the IR OA share.

*IR-6: Hypothesis on the aggregation of OA publications included in SR*

$H_1$: The larger the SR OA share of a university, the larger is the IR OA share.

$H_0$: The larger the SR OA share, the smaller (or equal) is the IR OA share.

---

[13]  Giesecke (2011, p. 531) calls this assumption the 'If you build it, they will come'-hypothesis.

[14]  See Picarra & Swan (2015); Lovett et al. (2017); and Herrmannova et al. (2019) for the compliance of researchers with OA policies.

[15]  In the statistical model, the variables that are summarised in this hypothesis are tested individually.

[16]  For the two aggregation hypotheses (IR-5 and IR-6), see the examples in Poynder (2016) and Tsay et al. (2017).

# 4. Data and methods

This study is based on three types of data: bibliometric data of the publication output complemented with OA evidence, structural data and information about OA infrastructures, and interviews with OA officers and OA representatives from German universities.

The first type of data determines the *publication output* of German universities and was retrieved from the Web of Science in-house database maintained by the German Competence Center for Bibliometrics (WoS-KB) in its 2021 version. The main advantage of this data source is that it provides disambiguated address information (Rimmert et al., 2017), which allows obtaining the publication output represented in the Web of Science on the level of institutions with a "near-complete national-scale coverage" of Germany's institutions at high accuracy (Donner et al., 2020). Given that both IR OA and SR OA can be achieved via self-archiving and given that all authors of a publication can in principle perform such activity, the publication output of a university was defined as publications with at least one author with an address from the particular German university. All publications of the period 2010-2020 with an author from a German university were considered. To identify items in repositories, article-level evidence from Unpaywall was used.[17] An article was classified as repository-based OA, if Unpaywall's field 'host type' reported that a version of the article was in a repository. Given that Unpaywall does not further distinguish between different types of repositories, domains from repository full text links were extracted from Unpaywall and matched with the Directory of Open Access Repositories (OpenDOAR), a comprehensive registry of repositories supporting the OAI standard. Using the OpenDOAR repository classification, we distinguished between institutional, discipline-based, and other types of repositories. If a domain was not listed in OpenDOAR, repository full-texts were classified as "other". For the further analysis, only articles in IR and SR were considered.

*The disciplinary profile of universities* was conceptualised by one factor on a high level of aggregation. For each of the 255 WoS subject categories, subject category-specific OA shares were calculated including all publications with at least one author address from a German institution. Based on the subject category OA shares and the number of publications in each subject category, a disciplinary influence factor was calculated for all universities and for both SR OA ($X_1^{SR}$) and IR OA ($X_1^{IR}$), namely

$$X_1^{SR}(i) = \frac{1}{T_i}\sum_{s \in S}(N_{i,s} * P_s^{SR}), \text{ and } X_1^{IR}(i) = \frac{1}{T_i}\sum_{s \in S}(N_{i,s} * P_s^{IR})$$

where

$X_1^{SR}(i)$    SR OA disciplinary influence factor for university $i \in I$ the set of all included universities,

$X_1^{IR}(i)$    IR OA disciplinary influence factor for university $i \in I$,

$N_{i,s}$    Number of publications of university $i \in I$ in WoS subject category $s \in S$ the set of all WoS subject categories,

$T_i$    Total number of publications of university $i \in I$,

$P_s^{SR}$    SR OA share of WoS subject category $s \in S$, and

---

[17]    The method for the addition of OA evidence is described in detail in Hobert et al. (2021). For the documentation of code and data, see also https://doi.org/10.5281/zenodo.3892951 (accessed August 15, 2023).

$P_s^{IR}$          IR OA share of WoS subject category $s \in S$.

Given that regression analyses were performed for three different periods (2010-2020, 2017-2018 and 2020), disciplinary influence factors were calculated for each period.

For *local infrastructures and services,* a structured data collection was created[18] by harvesting different sources of information and by manual online search. The data set includes information about the size of universities (in terms of students, staff, professors, budget, and third-party funds[19]) as well as OA infrastructures and services that are provided on the local level. The last-mentioned data include information about the existence of IR, OA policies[20], OA officers[21], OA websites and OA activities like information events or workshops announced on the universities' websites.[22] Data collection took place between August and October 2021. The data are modelled as response variables $X_2$ to $X_8$ (see Table 1).

For the operationalisation of the third factor, *transformative agreements*, data are available for the agreements negotiated between project DEAL and the large publishing houses Wiley and Springer. Although these contracts operate on an all-in principle and include all German universities, the number of publications covered by the two DEAL contracts varies from university to university as their publication output in journals covered by the contracts differs. These are exactly the publications that possibly compete with deposition in a SR and, at the same time, can be used for aggregation in institutional repositories. For our analysis of a possible influence of the DEAL contracts, the publication year 2020 is considered as this is the only year for which the transformative agreements with Springer and Wiley both have been effective for the whole year and for which data are available.[23] For each university we calculated the share of the publication output covered by DEAL contracts as

$$X_9^D(i) = \frac{D_{(i)}}{TP_{(i)}}$$

where

$X_9^D(i)$    Share of OA publications covered by DEAL contracts for university $i \in I$, the set of all included universities,

$D_i$       Number of OA publications covered by DEAL contracts for university $i \in I$,

$T_i$       Total number of publications of university $i \in I$.

Table 1 gives an overview of the explanatory variables together with their labels that are considered in the regression models.

---

**Table 1: Variables in the regression models**

| Variable description | Label |
|---|---|
| Estimated SR OA share / IR OA share based on the composition of subjects | $X_1^{SR}$ / $X_1^{IR}$ |
| Existence of an Institutional repository | $X_2$ |
| Existence of an OA officer | $X_3$ |
| Existence of a webpage with OA information | $X_4$ |
| Existence of a webpage with information about OA activities | $X_5$ |
| Existence of a webpage with OA rights information | $X_6$ |
| Existence of an OA policy | $X_7$ |
| Months of OA policy adoption | $X_8$ |
| Share of journal articles covered by DEAL contracts | $X_9^D$ |
| Subject/Institutional repository OA share | $Y^{SR}$ / $Y^{IR}$ |

In order to put our statistical models into a broader context and to gain more detailed insights into how the different aspects of the three factors influence the OA shares on the level of organisations, expert interviews with OA officers and representatives from 20 different universities were conducted between February and June 2021. The selection of interviewees aims to represent a large diversity of perspectives and follows the selection scheme of maximum variation (Collins et al., 2006, p. 84). It includes interviewees from large and small universities, universities with strong or weaker OA adaption as well as universities with different disciplinary profiles (with and without a medical faculty, technical universities and universities with a broad disciplinary mixture). The interview guideline covers all factors that were included in the regression models, and the duration of the interviews varied between 47 and 119 minutes. For the analysis, all interviews were completely transcribed by a transcription service to guarantee maximum quality. In the course of the content analysis (Mayring, 2015) of the interviews, MAXQDA 2018 data analysis software was used, and a code tree was developed that consists of 166 codes with 3118 coded paragraphs assigned to them.

## Descriptive statistics

In a first step, descriptive statistics are reported for categorical and metrical explanatory and response variables. Note that the availability of data differs. Structural information about the German university landscape and about OA infrastructures were collected at a specific point in time when the manual research took place. In contrast, publication-based information like publication output, OA shares, and disciplinary influence scores can be calculated from data spanning different periods. Finally, information about publications covered by DEAL contracts is available for the publication year 2020 only. With the exception of DEAL-shares, publication-based indicators are given for three periods (2010-2020, 2017-2018, and 2020) for which regression models are calculated. The rationale for the selection of the three periods is to analyse and compare the influence of the types of factors for the whole 11-year period, the most recent period before the introduction of the DEAL contracts (2017-2018) and the period for which information about the DEAL contracts are available (2020).

Table 2 gives an overview of the descriptive statistics for categorical independent variables and illustrates that German universities differ regarding the mechanisms and activities they have implemented to support OA. While more than 80% of the universities have a website with OA information, about three quarters have an OA officer and an IR and nearly two thirds an OA policy. Only half of the universities provide OA rights information and a bit more than a third of them information about OA courses and training on their websites.

**Table 2: Descriptive statistics for categorical independent variables**

| Variable | True | True (%) | False | False (%) |
|---|---|---|---|---|
| $X_2$ (IR) | 78 | 75.0 | 26 | 25,0 |
| $X_3$ (OA officer) | 77 | 74.0 | 27 | 26.0 |
| $X_4$ (Webpage with OA information) | 87 | 83.7 | 17 | 16.4 |
| $X_5$ (OA activities) | 37 | 35.6 | 67 | 64.4 |
| $X_6$ (OA rights information) | 51 | 49.0 | 53 | 51.0 |
| $X_7$ (OA policy) | 67 | 64.4 | 37 | 35.6 |

Descriptive statistics for the duration since the adoption of OA policies at German universities are given in table 3. The first line includes both universities with and without OA policies. For universities without OA policy, the duration of policy adoption was defined as 0 months. The statistics in the second line are limited to universities with OA policies.

**Table 3: Descriptive statistics for metrical independent variables**

| Variable | Obser-vations | Mean | Std. Dev. | Min | Max |
|---|---|---|---|---|---|
| $X_8$ (Months of policy adoption, all universities) | 104 | 48.77 | 53.02 | 0 | 179 |
| $X_8$ (Months of policy adoption, universities with OA policy) | 67 | 75.70 | 48.14 | 0 | 179 |

The publication-based descriptive statistics are presented in table 4. The table includes descriptive statistics for the total number of publications, SR and IR OA shares as well as the SR and IR OA disciplinary influence factors for all periods. For the publication year 2020, a DEAL influence factor was calculated. In a first step, all indicators were calculated for each university that exceeded the threshold value of a publication output of 50 publications for the particular period. The threshold value was introduced to exclude distortions of the OA shares and disciplinary influence factors due to small publication output. In a second step, mean value and standard deviation were calculated and minimum and maximum values were given for the German university landscape. The results in the table show that all OA shares have increased for more recent years and for both repository-based OA types. The growth of the share for both types happens roughly at the same scale while the standard deviation is larger in the case of SR OA when compared with IR OA.

**Table 4: Publication-based indicators (independent and dependent variables)**

| Variable | Obser-vations | Mean | Std. Dev. | Min | Max |
|---|---|---|---|---|---|
| **Period 2010-2020** | | | | | |
| Total Publications* | 100 | 12,678.34 | 14,033.15 | 72 | 59,115 |
| $Y^{SR}$, SR OA share* | 100 | 23.75 | 10.41 | 1.19 | 43.08 |
| $Y^{IR}$, IR OA share* | 100 | 18.56 | 8.00 | 5.51 | 59.77 |
| $X_1^{SR}$, SR OA disciplinary influence factor* | 100 | 23.38 | 6.63 | 7.28 | 38.38 |
| $X_1^{IR}$, IR OA disciplinary influence factor * | 100 | 18.60 | 1.92 | 14.45 | 24.03 |
| **Period 2017-2018** | | | | | |
| Total Publications** | 88 | 2,848.36 | 2,809.43 | 52 | 11,553 |
| $Y^{SR}$, SR OA share** | 88 | 28.09 | 10.67 | 0 | 45.29 |
| $Y^{IR}$, IR OA share** | 88 | 23.45 | 8.42 | 8.07 | 59.33 |
| $X_1^{SR}$, SR OA disciplinary influence factor ** | 88 | 27.02 | 6.98 | 8.06 | 40.41 |
| $X_1^{IR}$, IR OA disciplinary influence factor ** | 88 | 23.68 | 2.33 | 18.79 | 29.86 |
| **Period 2020** | | | | | |
| Total Publications*** | 84 | 1,694.05 | 1,590.93 | 56 | 6,511 |
| $Y^{SR}$, SR OA share*** | 84 | 35.92 | 11.13 | 6.35 | 52.11 |
| $Y^{IR}$, IR, OA share*** | 84 | 25.86 | 10.05 | 10.40 | 62.05 |
| $X_1^{SR}$, SR OA disciplinary influence factor*** | 84 | 33.39 | 7.58 | 12.12 | 45.32 |
| $X_1^{IR}$, IR OA disciplinary influence factor*** | 84 | 24.19 | 1.90 | 19.94 | 28.35 |

*Universities with a publication output > 50 in 2010-2020
** Universities with a publication output > 50 in 2017-2018
*** Universities with a publication output > 50 in 2020

## Regression models

We now turn to the results of the regression models. Multiple linear regression analysis is an important statistical tool to test assumptions about structures and relations in data (Freedman, 2009). In regression analysis, the output variable is named dependent variable, and the variables that are assumed to have effects on the dependent variable are called independent variables. In our analysis, we calculate separate regression models for three time periods and two dependent variables each – the SR OA share and the IR OA share. Given that collinearity is problematic for regression analysis, variance inflation factors (VIF) were calculated for both OA types and all three periods using the STATA 11 *VIF* function.

**Table 5: Variance inflation factors**

| | $X_1^{SR}$ | $X_1^{IR}$ | $X_2$ | $X_3$ | $X_4$ | $X_5$ | $X_6$ | $X_7$ | $X_8$ | $X_9^{D}$ | $Y^{SR}$ |
|---|---|---|---|---|---|---|---|---|---|---|---|
| **Subject Repository OA** | | | | | | | | | | | |
| 2010-2020 | 1.25 | -- | 1.43 | 1.83 | 2.39 | 1.26 | 1.44 | 2.66 | 2.19 | -- | -- |
| 2017-2018 | 1.25 | -- | 1.61 | 1.58 | 2.37 | 1.20 | 1.29 | 2.41 | 1.99 | -- | -- |
| 2020 | | -- | 1.64 | 1.62 | 2.08 | 1.21 | 1.32 | 2.29 | 1.96 | 1.62 | -- |
| **Institutional Repository OA** | | | | | | | | | | | |
| 2010-2020 | -- | 1.77 | 1.56 | 1.83 | 2.39 | 1.28 | 1.556 | 2.74 | 2.24 | -- | 1.71 |
| 2017-2018 | -- | 2.26 | 1.72 | 1.56 | 2.37 | 1.21 | 1.36 | 2.50 | 2.05 | -- | 2.18 |
| 2020 | -- | 1.49 | 1.75 | 1.62 | 2.02 | 1.21 | 1.35 | 2.28 | 1.98 | 1.60 | 1.53 |

The values in Table 5 show that there is some explanatory power between the independent variables, but they all are well below the critical value of 5 that is considered as a threshold value above which the model should be adjusted, e.g. by excluding or modifying certain independent variables. As a consequence, all considered variables are included in the regression analysis.

**Table 6: Subject Repository OA, regression models**

| Reg. | $X_1^{SR}$ | $X_2$ | $X_3$ | $X_4$ | $X_5$ | $X_6$ | $X_7$ | $X_8$ | $X_9$ | F | $R^2$ | Adj. $R^2$ | RMSE |
|---|---|---|---|---|---|---|---|---|---|---|---|---|---|
| **2010-2020** | | | | | | | | | | | | | |
| 1 | 1.500** | 0.001 | -0.010 | -0.001 | 0.009 | 0.006 | -0.009 | 0.000 | -- | 171.39 | 0.938 | 0.923 | 0.027 |
| 2 | 1.515** | -- | -- | -- | -- | -- | -- | -- | -- | 1346.33 | 0.932 | 0.932 | 0.027 |
| **2017-2018** | | | | | | | | | | | | | |
| 3 | 1.437** | 0.010 | -0.008 | 0.001 | 0.010 | 0.003 | -0.011 | 0.000 | -- | 124.45 | 0.927 | 0.919 | 0.030 |
| 4 | 1.463** | -- | -- | -- | -- | -- | -- | -- | -- | 952.14 | 0.917 | 0.916 | 0.031 |
| **2020** | | | | | | | | | | | | | |
| 5 | 1.495** | 0.013 | -0.016 | -0.034 | 0.008 | 0.003 | 0.010 | 0.000 | -0.000 | 116.23 | 0.934 | 0.926 | 0.030 |
| 6 | 1.411** | -- | -- | -- | -- | -- | -- | | | 1017.14 | 0.925 | 0.924 | 0.031 |

* Significant at the 0.05 level; ** significant at the 0.01 level

**Table 7: Institutional Repository OA, regression models**

| Reg. | $X_1^{IR}$ | $X_2$ | $X_3$ | $X_4$ | $X_5$ | $X_6$ | $X_7$ | $X_8$ | $X_9$ | $Y^{SR}$ | F | $R^2$ | Adj. $R^2$ | RMSE |
|---|---|---|---|---|---|---|---|---|---|---|---|---|---|---|
| **2010-2020** | | | | | | | | | | | | | | |
| 7 | 1.885** | 0.006 | -0.001 | 0.010 | 0.025 | -0.015 | 0.014 | 0.000 | -- | 0.231* | 11.01 | 0.524 | 0.476 | 0.058 |
| 8 | 1.937** | -- | -- | -- | -- | -- | -- | -- | -- | 0.252** | 47.07 | 0.493 | 0.482 | 0.058 |
| 9 | 2.693** | -- | -- | -- | -- | -- | -- | -- | -- | -- | 70.31 | 0.418 | 0.412 | 0.061 |
| **2017-2018** | | | | | | | | | | | | | | |
| 10 | 1.998** | 0.004 | -0.002 | -0.008 | 0.026 | -0.009 | 0.001 | 0.000 | -- | 0.136 | 10.31 | 0.543 | 0.491 | 0.060 |
| 11 | 2.538** | -- | -- | -- | -- | -- | -- | -- | -- | -- | 84.40 | 0.495 | 0.489 | 0.060 |
| **2020** | | | | | | | | | | | | | | |
| 12 | 1.762** | 0.005 | 0.019 | 0.015 | 0.023 | -0.026 | 0.010 | 0.000 | 0.000 | 0.115 | 2.02 | 0.217 | 0.110 | 0.095 |
| 13 | 1.988** | -- | -- | -- | -- | -- | -- | -- | -- | -- | 13.46 | 0.141 | 0.131 | 0.094 |

* Significant at the 0.05 level; ** significant at the 0.01 level

A first look at the results of the regression analyses reveals that the uptakes of the two types of repository-based OA follow different patterns. To begin with subject repository OA, the only factor that is significant and that has strong explanatory power for each of the three periods is the disciplinary influence factor. According to Adj. $R^2$ for the period 2010-2020, it explains 93.2% of the variance of the subject repository OA share of German universities, and for more recent periods the share is on a similar level (91.6% for 2017-2018 and 92.4% in 2020). For SR-1 ("Universities with a disciplinary profile that shows a strong affinity towards SR OA have a larger SR OA share than universities with a weaker disciplinary affinity towards SR OA"), the null-hypothesis is therefore rejected, meaning that the disciplinary profile of a university has a strong influence on the SR OA share. Regarding the variables that operationalise local infrastructures and services, none of them turned out to be significantly different from zero. The respective null-hypotheses of SR-2 (where $H_1$: "Universities with an OA policy have a larger SR OA share than universities without an OA policy") and SR-3 (where $H_1$: "Universities with an OA officer, websites with OA (rights) information and/or training activities have a larger SR OA share than universities without such OA-supporting provisions") cannot be rejected. In other words, the existence of local infrastructures and services does not affect the subject repository OA share. The influence of transformative agreements could only be tested for the period 2020. SR-4 ($H\_1$: "The larger the share of the publication output covered by transformative agreements, the smaller is the subject repository OA share of German universities") suggests a possible de-incentivising effect for the deposition of articles on subject repositories which may result from the fact that such publications are already OA in the publisher version. Given that the coefficient of the DEAL influence factor is not significantly different from 0, the null-hypothesis cannot be rejected, casting doubt on the relevance of such an effect. To summarise, the regression models provide evidence that differences in the SR OA share of German universities result from the composition of the disciplinary profile with their specific OA affinity, but there is no evidence for an influence of any of the other variables that are considered here.

Compared with SR OA, the regression models for IR OA are somewhat more heterogeneous and a bit more difficult to interpret: for all of the three periods, the disciplinary influence factor is again the independent variable with the strongest explanatory power and it is also highly significant. As a result, the null-hypothesis of IR-1 ($H_0$: "Universities with a disciplinary profile that shows a strong affinity towards IR OA have a smaller (or equal) IR OA share than universities with a weaker disciplinary affinity towards IR OA") is rejected, implying a strong influence of the disciplinary profile. However, two additional aspects are worth noticing. First, the variance explained by the disciplinary influence factor is considerably lower for IR OA (reg. 9, 11, 13) than for SR OA (reg. 2, 4, 6). Second, the explained variance is large for the period 2010-2020 (adj. $R^2$: 0.412) and 2017-2018 (adj. $R^2$: 0.489) and is strongly shrinking in 2020 (adj. $R^2$: 0.131). These results point to a decreasing relevance of the disciplinary profile as an explanation of the IR OA share for the last year that is considered in the analysis.

With respect to IR-2 (where $H\_1$: "Universities that provide an IR for self-archiving have a larger IR OA share than universities that do not provide such a repository"), the regression coefficients for $X_2$ are not significantly different from 0 and the null-hypothesis can not be rejected. An influence of the provision of an IR is therefore highly improbable. The result that the mere existence of an IR is not sufficient to provoke self-archiving is in line with Giesecke (2011). Again, the regression coefficients for $X_7$ and $X_8$ are not significantly different from 0 in regressions no. 7, 10 and 12, indicating that universities with an OA policy do not have larger IR OA shares than universities without such OA supporting instruments: the null-hypothesis of IR-3 (where $H_1$:"Universities with OA policy have a larger IR OA share than universities that do not have such a policy") cannot be rejected. The same holds for the results for IR-4 (where H_1: "Universities with an OA officer, websites with OA (rights) information and/or OA training activities and events have a larger IR OA share than universities without such OA-supporting provisions"). In all periods analysed here, the variables $X_4$ to $X_8$ turned out not to be significantly different from 0. Regarding the two remaining hypotheses on the effects of aggregating activities at libraries (IR-5, where $H_1$: "The larger the share of the publication output of a university covered by transformative agreements, the larger is the IR OA share" and IR-6, where $H_1$: "The larger the SR OA share, the larger is the IR OA share (as university libraries find more OA publications that can be aggregated in their IR)"), neither of the null-hypotheses can be rejected for any of the periods with one exception: for the period 2010-2020, the coefficient of the SR OA share $Y_S$ is significantly different from 0 on a 0.01-level adding 7.0% explained variance to the model. With tolerable probability of error, a higher availability of publications on subject repositories seems to increase the IR OA share in the period 2010-2020.

# 6. Discussion

In this step, the results of the regression analyses are discussed in the context of findings in the literature as well as results of interviews with 20 OA administrators from German universities that were conducted within this study.

## Subject Repositories

To begin with subject repository OA, the results show that the disciplinary profile is the most determining factor for the understanding of differences in this OA type. In other words, the adoption of SR OA happens in the first place within certain disciplines and specialties, and the SR OA share of an institution reflects the disciplinary profile and its composition with fields that have more or less affinity towards SR OA. This finding is in line with the observations of the large majority of the interviewees that report large differences between different disciplines in the extent to which self-archiving on SR takes place. One example is interviewee I-05 who describes the situation of such differences between disciplines as follows:

> "Given that it has developed in principle in high energy physics, actually since the internet came into existence and has spread a little in neighbouring disciplines, takes all physics and mathematics on top. You can see that they were used to doing such things. And it is the same with economics. There have always been discussion papers that circulated, where preprints were established and this has been transferred into the electronic world. In other cases, such predecessors are missing to some extent". (I-05, pos. 53)[24]

---

[24]    The interviews were conducted in German. All quotations of the interviews are translations by the authors.

Other interviewees also point to long-standing traditions in the publication culture of certain fields with stable patterns of exchange of preprints and working papers.[25] Mechanisms stabilising such type of exchange that were addressed in the interviews are the combination of a better findability and accessibility for scientists in the role of the reader[26] and better visibility of their own research and a means to increase reputation for scientists in the role of the author.[27] Such complementary patterns in the use of preprint servers are involved in some (Taubert, 2021) but not in all disciplines, as studies on the usage of SR show (Kling & McKim, 2000; Björk et al., 2014; Pinfield et al., 2014). An additional argument put forward by two interviewees that may help to explain disciplinary differences in the adoption of SR OA is that the deposition of preprints on SR is also a means for speeding up the circulation of new findings (I-07, pos.9, I-15, pos. 71). However, speed of circulation may not be of similar relevance in all disciplines and fields.

With respect to SR, one interesting aspect is that a number of interviewees[28] report that they do not know much about the self-archiving on SR of scientists from their institution and often do not regard themselves as being competent to answer the questions on SR. The reason for this becomes visible in the following quotation from the interview with I-16:

> "I don't know much about it. By coincidence I notice that some papers that are funded by our publication fund have been self-archived by natural scientists on arXiv. This is what I notice sporadically. Well, regarding the social sciences and humanities, I do not notice anything because we do not have a working bibliography for our university and because they rarely use our publication fund. Therefore, I cannot say much about it". (I-16, pos. 43)

OA administrators are often not aware of the self-archiving activities from scientists of their university as they are responsible for a certain set of OA services like the provision of a publication fund or the operations of a publication platform or an IR. A number of interviews suggest that the responsibilities of OA administrators are not directed towards the increase of OA based on SR in the first place and that they only get in contact with this OA type when native OA services of the university – like publication funds in the quotation above – are involved. This orientation towards own native services may explain in part why the variables that operationalise local infrastructures and services do not play a role for the explanation of the differences of the SR OA share, as they do not focus on this OA type at many universities.

## Institutional Repository OA.

Compared with SR, the results of the regression models for IR are less outstanding, but a solid share of the variance of the IR OA share is explained for the two periods 2010-2020 and 2017-2018 by the disciplinary influence factor. In 2020, there is a remarkable decrease down to 13.1% of explained variance. Again, the variable of local infrastructures and services do not add explained variance to the model while the 'SR OA share' adds 7.0% of explained variance for the period 2010-2020 only. These results provoke two questions: first, what are the reasons for the substantial decrease of the explanatory power of the regression model in the most recent year? Second, why is the explanatory power of the variables of local infrastructures and services that low?

---

[25]  I-04, pos. 22; I-05, pos. 51; I-07, pos. 7, I-12, pos. 37, I-02, pos. 70-72, I-13, pos. 25; pos. 27, I-14, pos. 27, I-19, pos. 14.

[26]  I-07, pos.9.

[27]  I-01, pos. 53; I-08, pos. 42; I-19, pos. 60, pos. 66; I-20, pos. 33, pos. 117.

[28]  I-04, pos. 38, I-09, pos. 49; I-11, pos. 29, I-12, pos. 99, I-15, pos. 71, I-16. Pos. 43 I-17, pos. 51, I-20, pos. 33.

Regarding the first question, the results of the regression analysis for 2020 suggest that the main factors for this period are not included in the model and that a change happened between the period 2017-2018 and 2020. Fortunately, the interviews provide some evidence of a factor that became relevant in recent years. To understand this change, it is key to note that the diagnosis of a low usage that has accompanied IR for more than 15 years (Westrienen & Lynch, 2005; Xia, 2008; Nicholas et al., 2012; Arlitsch & Grant, 2018; Novak & Day, 2018) is also reflected in the interviews conducted for this study. No less than 13 interviewees[29] state phrases like the amount of self-archiving is 'very little' (I-04, pos. 30), 'very, very, very low' (I-20, pos. 25), or 'could definitely be more' (I-15, pos. 49), and that the activities of the scientists on IR are below their expectations. Two responses to low usage can be found in the interviews. One is to provide 'value-added services' to incentivise self-archiving. Examples are the possibility to create personal or project-specific publication lists for different purposes[30] or the integration of the Open Researcher and Contributor ID (ORCID) and metadata exchange with this service. The other one consists of a large diversity of activities of libraries that aim to collect or aggregate publications in the IR by researchers from their institution and can be summarised under the term 'mediated archiving' (Xia 2008). Such activities often happen autonomously by the operators of the IR without any consent of the authors being necessary. In 14[31] of the 20 interviews, such activities were characterised as a strategy besides self-archiving to fill repositories with content. Sources for full texts to be deposited by libraries in IR are diverse and can be assigned to two types: on the one hand, there are non-OA sources. In these cases, archiving on IR is a means to make a publication OA. Examples are the project DeepGreen with its original aim to operate as a data hub that automatically collects full texts and metadata of publications from publishers for which secondary publication is permitted by, for example, national licences to operators of IR for automatic depositing (Boltze et al., 2022). According to the project website,[32] 49 German universities used this service at the time of writing, and it is also mentioned as a source of content for IR in a number of interviews.[33] Universities also use their subscription licences that include self-archiving rights for the institutions' authors to deposit their publications on the IR. On the other hand, the interviewees mention a number of sources for aggregation where the content is already OA. These include SR[34], publications in full OA journals paid by the university's publication fund[35], publications in journals of OA publishers[36], as well as publications made OA by transformative

---

agreements.[37] The use of such data sources can be understood as a remarkable shift of the original aim of IR to increase OA to the universities' intellectual output. One interviewee describes this development as follows:

> "Well what we do is, we put everything on [Name of IR, anonymized] what is funded via our publication fund or via DEAL or via other transformative agreements. But this is redundant because these are primary Gold OA publications. They can also be found on [Name of IR, anonymized] and we can say 'Look at all the nice things we have funded'. But it is a type of secondary publishing or archiving but not in a way that something becomes accessible that was formerly behind a paywall". (I-16, pos. 33)

In this context of use, the deposition of content that is already OA on IR can be understood at best as a strategy of archiving that guarantees OA articles to stay OA in a publishing environment which bears the risk that certain journals may vanish (Laakso et al. 2021) or as an attempt to create legitimation for a repository that is rarely used by researchers. The re-definition of the role of IR, however, does not stop here and reaches beyond the deposition of full texts. A number of interviewees[38] report that IR are used for the aggregation of metadata. In this context, the designated use of IR has shifted from an infrastructure that aims to support the supply of scientific information to the information requirements of the universities' management.

Coming back to the first question about the decreasing explanatory power of the regression models for more recent years, evidence from the interviews support the assumption that important factors are not considered in the models. These may include the myriad ways in which university libraries collect and aggregate content from manifold sources in their IR. Many of these sources originated and became usable during the last few years, such as DeepGreen, which disseminated its service from 2018 onwards, the transformative agreements with project DEAL, which became effective from 2020 onwards, or also the publication funds that were created during the last years at many universities.

Regarding the second question about the low explanatory power of the variables of local infrastructures and services, the individual variables have been discussed elsewhere in detail (Taubert et al. 2023) and shall not be repeated here. However, we would like to summarise the most important aspects. First, the non-significance of OA policies in the regression model is in line with the discussion of compliance rates in the literature, where high compliance is only reported for strong mandates (OA is required) linked to sanctions, while OA policies, like in the case of Germany, request OA and result in low compliance (Gargouri et al., 2012; Vincent-Lamarre et al., 2016; Larivière & Sugimoto, 2018; Kirkmann & Haddow, 2020). Second, the interviews illustrate that the collection and aggregation of content in IR is an important field of activity of university libraries and may influence the IR OA share. However, the data sources that are used and the aggregation strategies differ. Against the background of the interviews, the mere existence of a repository, an OA administrator, a website with OA (legal) information or OA training activities does not explain much and seems to be an oversimplified operationalisation of the diverse manifestation of infrastructures and services that are provided by universities. For further research about the determinants of the IR OA share, it is probably more suitable to ask for the collection and aggregation of activities undertaken by libraries, the data sources that are used and the manpower that is applied for this task than for supporting infrastructures and services that refer to self-archiving.

---

[37]  In the German context most prominently (but not only) the contracts of project DEAL (mentioned in Interviews I-05, pos. 41; I-07, pos. 21; I-15, pos. 49; I-16, pos. 33; I-14, pos. 33).

[38]  I-01, pos. 45; I-02, pos. 54; I-05, pos. 45; I-07, pos. 19, pos. 23; I-11, pos. 27; I-13; pos. 19.

# 7. Conclusion

The aim of this article is to answer the question as to what extent the disciplinary profile, local infrastructures and services and transformative agreements of the project DEAL determine the repository-provided OA share of German universities. The results show that the two repository OA types, subject repository OA (SR OA) and institutional repository OA (IR OA), follow different logics. Therefore, the question has to be answered separately for each of them. Regarding the SR OA share of universities, the regression analyses convincingly show that the composition of the disciplinary profile with a stronger or lesser affinity towards OA is decisive and explains as much as 93.2% of the variance for the period 2010-2020. For more recent years, the importance of the disciplinary profile is on the same level with an explained variance of 91.6% for 2017-2018 and 92.4% for 2020. This result highlights that the adoption of SR is in the first place driven by the inner logic of scientific publication cultures, which is in this context referred to as the disciplinary profile. All other factors like infrastructural support do not help to understand differences of the SR OA share of institutions.

With respect to the IR OA share, the composition of the disciplinary profile is again the most determining factor, but the explained variance is lower than in the case of IR OA. For the period 2010-2020, the disciplinary influence factor explains 41.2%, and for the period 2017-2018, 48.9% of the variance of the universities IR OA share. Variables that operationalise local infrastructure and services are all non-significant and do not improve the regression models. For 2020, the decrease of the variance explained by the disciplinary profile of the university to 13.1% suggests that a change is taking place regarding the determining factors for universities´ IR OA share.

Evidence from qualitative interviews with 20 OA administrators from German universities show that the designated use of IR has shifted in the past. Besides the collection of metadata that support the information needs of the universities' management, university libraries have become active in the aggregation of content from other sources. Such activities could be a relevant factor in the explanation of the IR OA share. Aggregation thereby does not refer to articles behind paywalls only with the objective to make them OA – like in the case of DeepGreen – but also to content that is already OA. At best such usage of IR can be understood as a strategy of archiving that guarantees OA articles to remain OA or as an attempt to create legitimation for a repository that is rarely used by researchers. Further research is well advised to consider such collection and aggregation activities that university libraries undertake when aiming to explain differences in the IR OA share of universities.